# Double beta decay of $^{150}$Nd to the first excited $0^+$ level of $^{150}$Sm: preliminary results

A.S. Barabash[1], P. Belli[2,3], R. Bernabei[2,3], R.S. Boiko[4,5], F. Cappella[6], V. Caracciolo[7], R. Cerulli[2,3], F.A. Danevich[4], A. Di Marco[2,3], A. Incicchitti[6,8], D.V. Kasperovych[4,*], R.V. Kobychev[4], V.V. Kobychev[4], S.I. Konovalov[1], M. Laubenstein[7], D.V. Poda[4,9], O.G. Polischuk[4], V.I. Tretyak[4], V.I. Umatov[1]

[1] *National Research Centre Kurchatov Institute, Institute of Theoretical and Experimental Physics, 117218, Moscow, Russia*
[2] *INFN, sezione di Roma "Tor Vergata", I-00133, Rome, Italy*
[3] *Dipartimento di Fisica, Università di Roma "Tor Vergata", I-00133, Rome, Italy*
[4] *Institute for Nuclear Research, 03028 Kyiv, Ukraine*
[5] *National University of Life and Environmental Sciences of Ukraine, 03041 Kyiv, Ukraine*
[6] *INFN, sezione di Roma, I-00185 Rome, Italy*
[7] *INFN, Laboratori Nazionali del Gran Sasso, I-67100 Assergi (AQ), Italy*
[8] *Dipartimento di Fisica, Universita di Roma "La Sapienza", I-00185 Rome, Italy*
[9] *CSNSM, Univ. Paris-Sud, CNRS/IN2P3, Université Paris-Saclay, 91405 Orsay, France*

*Corresponding author: dkasper@kinr.kiev.ua

The double beta decay of $^{150}$Nd to the first excited $0^+$ level of $^{150}$Sm ($E_{exc}$ = 740.5 keV) has been investigated with the help of the ultra-low-background setup consisting of four HP Ge (high-purity germanium) detectors ($\simeq$ 225 cm$^3$ volume each one) at the Gran Sasso underground laboratory of INFN (Italy). A highly purified 2.381-kg sample of neodymium oxide (Nd$_2$O$_3$) was used as a source of γ quanta expected in the decays. Gamma quanta with energies 334.0 keV and 406.5 keV emitted after deexcitation of the $0_1^+$ 740.5 keV level of $^{150}$Sm are observed in the coincidence spectra accumulated over 16375 h. The half-life relatively to the two neutrino double beta decay $^{150}$Nd → $^{150}$Sm($0_1^+$) is measured as $T_{1/2} = [4.7^{+4.1}_{-1.9}(\text{stat}) \pm 0.5(\text{syst})] \times 10^{19}$ y, in agreement with results of previous experiments.

## 1. Introduction

The double beta (2β) decay is a spontaneous transformation of ($A$, $Z$) nucleus to ($A$, $Z \pm 2$), which can occur in two main modes. In the two neutrino (2ν) mode, allowed in the Standard Model of particle physics (SM), the emitted electrons are accompanied by two (anti)neutrinos. The 2ν2β decay, being a second-order process in perturbation theory, is the rarest process observed in nature with half-lives in the range $\sim 10^{18} - 10^{24}$ y [1 - 3]. In neutrinoless double beta decay (0ν2β) no neutrinos are expected. Therefore, this process is forbidden in the SM due to the lepton number violation by two units. Nevertheless, 0ν2β decay is predicted in many SM extensions [4 - 9] where the neutrino is expected to be a Majorana particle (neutrinos and antineutrinos are equal) with non-zero masses [10]. The evidence of a finite neutrino mass was obtained in many experiments where the effect of neutrino oscillations was observed (see [11] and references therein). While the oscillation experiments are sensitive to the squared neutrino mass eigenstates difference, investigations of 0ν2β decay is the only realistic way to determine the absolute neutrino mass scale and the neutrino mass hierarchy, to test the lepton number conservation, the nature of neutrino (Dirac or Majorana particle) and many other effects beyond the SM.

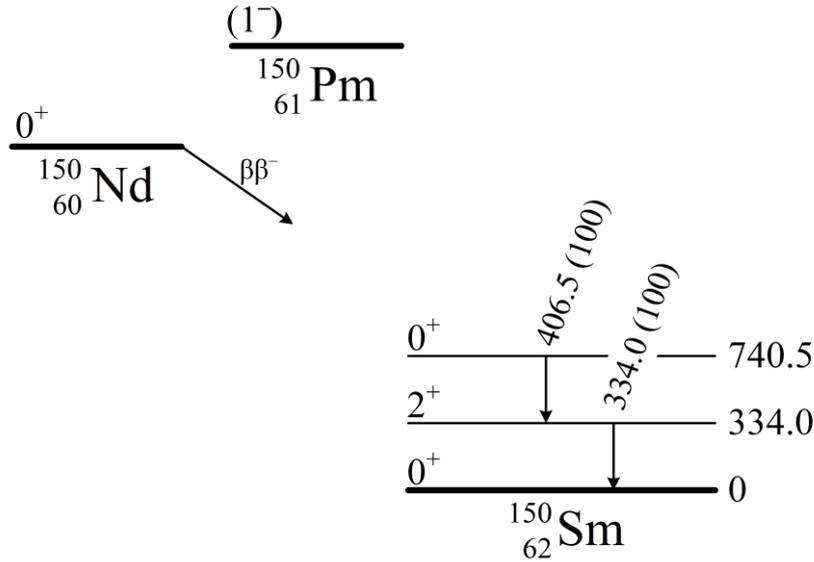

Fig. 1. A simplified decay scheme of $^{150}$Nd → $^{150}$Sm($0_1^+$) 2β decay [17]. The energies of the levels and of the emitted γ quanta are in keV (relative intensities of γ quanta are given in parentheses in %).

The nuclide $^{150}$Nd is one of the most promising among the 35 naturally occurring 2β$^-$ isotopes [1] thanks to the one of the highest energy release $Q_{\beta\beta} = 3371.38(20)$ keV [12] and a high natural isotopic abundance δ = 5.638(28) % [13]. The 2ν2β decay of $^{150}$Nd to the ground state of $^{150}$Sm (a simplified decay scheme of $^{150}$Nd is presented in Fig. 1) was measured in several direct experiments in the range of $T_{1/2} = (0.7 - 1.9) \times 10^{19}$ y [14 - 16].

In addition to the 2β decay of $^{150}$Nd to the ground state, the transition to the first $0^+$ 740.5 keV excited level of $^{150}$Sm was observed too with the half-life values $T_{1/2} = (7 - 14) \times 10^{19}$ y [18 - 21]. A summary of all the experiments where this specific decay was detected is given in Table 1.

Table 1. Summary of the investigations of the 2ν2β decay of $^{150}$Nd to the first $0^+$ 740.5 keV excited level of $^{150}$Sm. The statistical and systematic uncertainties of the $T_{1/2}$ values, given in the original papers, are added in squares. The result of the NEMO-3 experiment is not published yet and is given only as preliminary one.

| Short description | $T_{1/2}$, $10^{19}$ y | Year [Ref.] |
|---|---|---|
| Modane underground laboratory (4800 m w.e.), HP Ge 400 cm$^3$, 3046 g of Nd$_2$O$_3$ (δ = 5.638%), 11321 h, 1-d spectrum | $14^{+5}_{-4}$ | 2004 [18] |
| Re-estimation of the result [18] | $13.3^{+4.5}_{-2.6}$ | 2009 [19] |
| Modane underground laboratory (4800 m w.e.), NEMO-3 detector, foil with 57.2 g of $^{150}$Nd$_2$O$_3$ (δ = 91.0%), 40774 h, energies of e$^-$ and γ, tracks for e$^-$ (preliminary result) | $7.1 \pm 1.6$ | 2013 [20] |
| Kimballton Underground Research Facility, 2 HP Ge (~304 cm$^3$ each one), 50 g $^{150}$Nd$_2$O$_3$ (δ = 93.6%), 15427 h, coincidence spectrum | $10.7^{+4.6}_{-2.6}$ | 2014 [21] |
| Gran Sasso underground laboratory (3600 m w.e.), 4 HP Ge (~225 cm$^3$ each one), 2381 g of Nd$_2$O$_3$ (δ = 5.638%), 16375 h, sum of 1-d spectra, coincidence spectrum | $4.7^{+4.1}_{-1.9}$ | This work |

## 2. Experiment

### 2.1. Purification of Nd$_2$O$_3$

The sample of high purity Nd$_2$O$_3$, produced by a Soviet Union industry in the 70-s, utilized in previous experiment [18], was additionally purified by using combinations of chemical and physical methods [22, 23]. First, the neodymium oxide was dissolved in high purity hydrochloric acid:

$$Nd_2O_3 + 6HCl \rightarrow 2NdCl_3 + 3H_2O. \tag{1}$$

Partial precipitation from the acidic solution was obtained by increasing the pH level up to 6.5 - 7.0 with ammonia gas. The procedure was realized for co-precipitating of Th and Fe impurities, taking into account that hydroxides of these elements precipitated at a lower pH level than the neodymium oxide.

To realize liquid-liquid extraction, the solution was acidified with diluted hydrochloric acid down to pH ≈ 1. The liquid-liquid extraction method is based on extraction of compound from the solvent A to the solvent B, when A and B are not miscible. The neodymium chloride was dissolved in water (phase A), while a solution of phosphor-organic complexing compound tri-*n*-octyl-phosphine oxide (TOPO) in toluene was used as a solvent B.

Elements with a higher oxidation preferably move to the organic phase with a higher distribution level in comparison to elements with a lower oxidation. Thus, this method allows to separate elements with different oxidation states [24]. This process can be written as

$$NdCl_3(Th, U)_{(aq)} + nTOPO_{(org)} \rightarrow NdCl_{3(aq)} + [(Th,U) \cdot nTOPO](Cl)_{(org)}. \quad (2)$$

The liquids were mixed together over 5 min, then the solutions were completely stratified in 30 min. The purified $NdCl_3$ was separated using separatory funnel. The amorphous neodymium hydroxide was obtained from the solution by using gaseous ammonia:

$$NdCl_3 + 3NH_3 + 3H_2O \rightarrow Nd(OH)_3\downarrow + 3NH_4Cl. \quad (3)$$

The purified $Nd_2O_3$ was obtained from the hydroxides by high temperature decomposition:

$$2Nd(OH)_3 \xrightarrow{900^0 C} Nd_2O_3 + 3H_2O. \quad (4)$$

The yield of the purified material was ~90 %.

**2.2. Low-background measurements**

The experiment is carried out deep underground (~3600 m w.e.) at the STELLA facility of the Gran Sasso underground laboratory [25]. The $Nd_2O_3$ sample with a total mass 2.381 kg, pressed into 20 cylindrical tablets $56 \pm 1$ mm in diameter and $16 \pm 0.5$ mm of thickness, was installed in the GeMulti ultra-low-background HP Ge gamma-spectrometer with four germanium detectors with volumes of 225.2, 225.0, 225.0 and 220.7 $cm^3$. The detectors are assembled in a cryostat with a cylindrical well in the center. The detectors are shielded by radiopure copper (10 cm) and lead (20 cm). The whole setup is enclosed in a Plexiglas box flushed with high-purity nitrogen gas to remove radon.

The data acquisition system of the spectrometer records the time and the energy of the events occurring in each detector and it allows to study the coincidence between the detectors. The energy scale and resolution of the HP Ge detectors were measured at the beginning of the experiment with $^{22}$Na, $^{60}$Co, $^{133}$Ba, $^{137}$Cs and $^{228}$Th γ-sources. Then the individual spectra were transformed to the same energy scale by using background

gamma peaks with energies 609.3, 1120.3 and 1764.5 keV ($^{214}$Bi), 351.9 keV ($^{214}$Pb), 911.2 keV ($^{228}$Ac), 1460.8 keV ($^{40}$K) and 2614.5 keV ($^{208}$Tl) using the algorithm described in [26]. As a result, the gamma peaks positions in the cumulative spectrum deviate from their table values [27] by less than 0.2 keV. The final energy resolution in the cumulative spectrum gathered with the Nd$_2$O$_3$ sample over 16375 h can be described by the following function: FWHM=$\sqrt{2.7(5)+0.0025(5)\cdot E_\gamma}$, where FWHM and $E_\gamma$ (energy of γ quanta) are in keV.

The cumulative energy spectrum accumulated with the Nd$_2$O$_3$ sample over 16375 h is shown in Fig. 2 together with the background spectrum measured without samples during 7862 h [28].

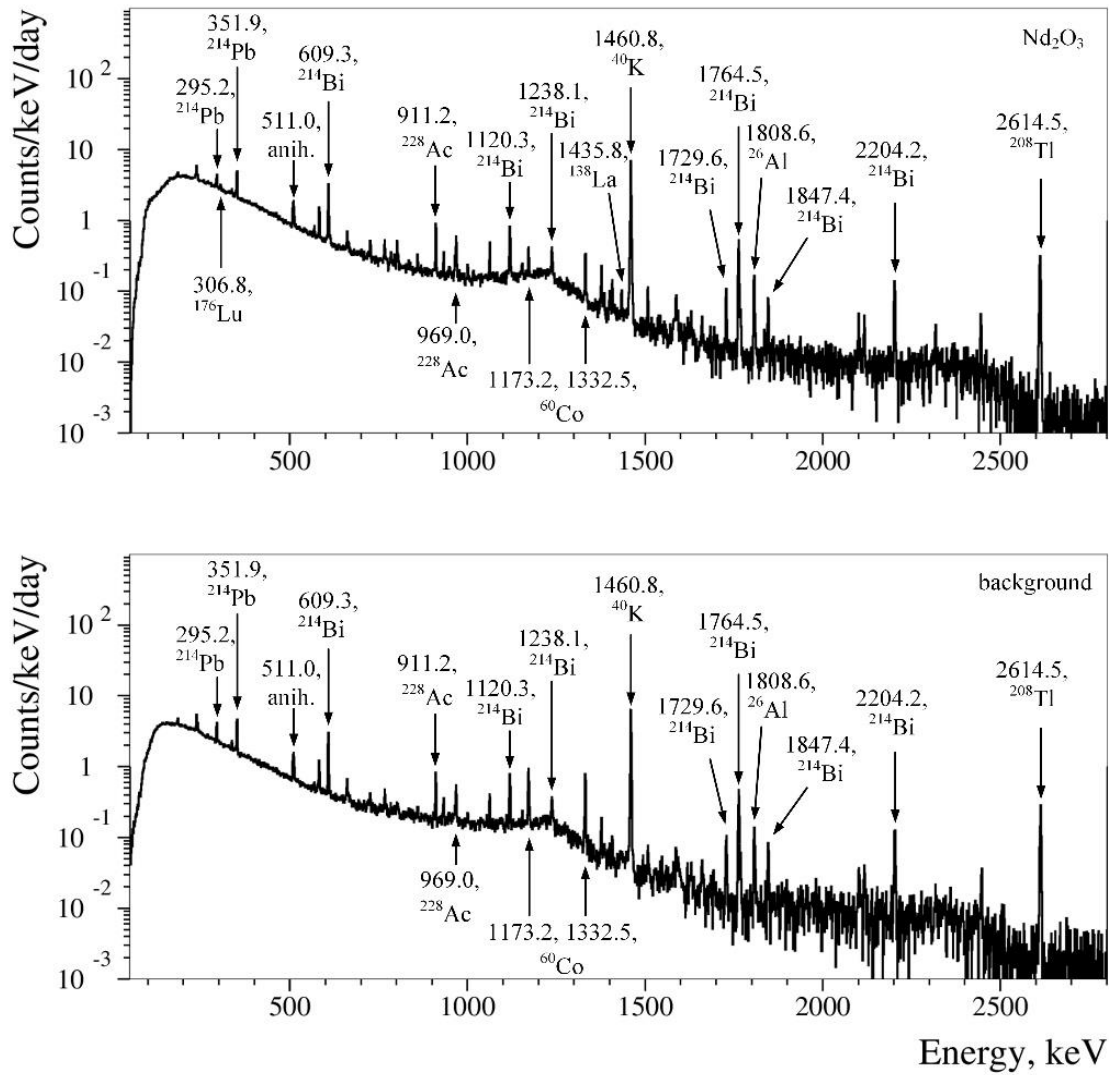

Fig. 2. The energy spectrum measured over 16375 h with the 2.381-kg Nd$_2$O$_3$ sample (*top*) and the background spectrum collected for 7862 h (*bottom*). Energies of gamma quanta are given in keV.

## 3. Results and discussion

### 3.1. Radioactive contaminations of the Nd$_2$O$_3$ sample

As it was described in Sec. 2.1, the neodymium oxide sample was purified to remove residual contamination of the material, particularly by potassium, radium and lutetium. The radioactive contaminations of the neodymium oxide before and after the purification were measured in the STELLA facility by using the ultra-low-background HP Ge detector GePaolo with a volume of 518 cm$^3$. The detector is shielded with radiopure copper (5 cm) and lead (25 cm). The whole setup is flushed by a high-purity nitrogen gas to remove radon and its progeny. The energy resolution of the spectrometer was about 2 keV for 1333 keV γ quanta of $^{60}$Co. The sample, sealed in a thin polyethylene film, was placed directly on the endcap of the detector. In both the spectra, measured with the Nd$_2$O$_3$ sample and in the background one, there are γ peaks that can be ascribed to $^{40}$K, $^{137}$Cs, $^{60}$Co, and radionuclides from the $^{238}$U and $^{232}$Th chains, while the gamma peaks at 1435.8 keV ($^{138}$La) and 306.8 keV ($^{176}$Lu) were observed only in the data accumulated with the Nd$_2$O$_3$ sample due to contamination of the material by lanthanum and lutetium. The estimation of radionuclides content in the Nd$_2$O$_3$ sample is summarized in Table 2.

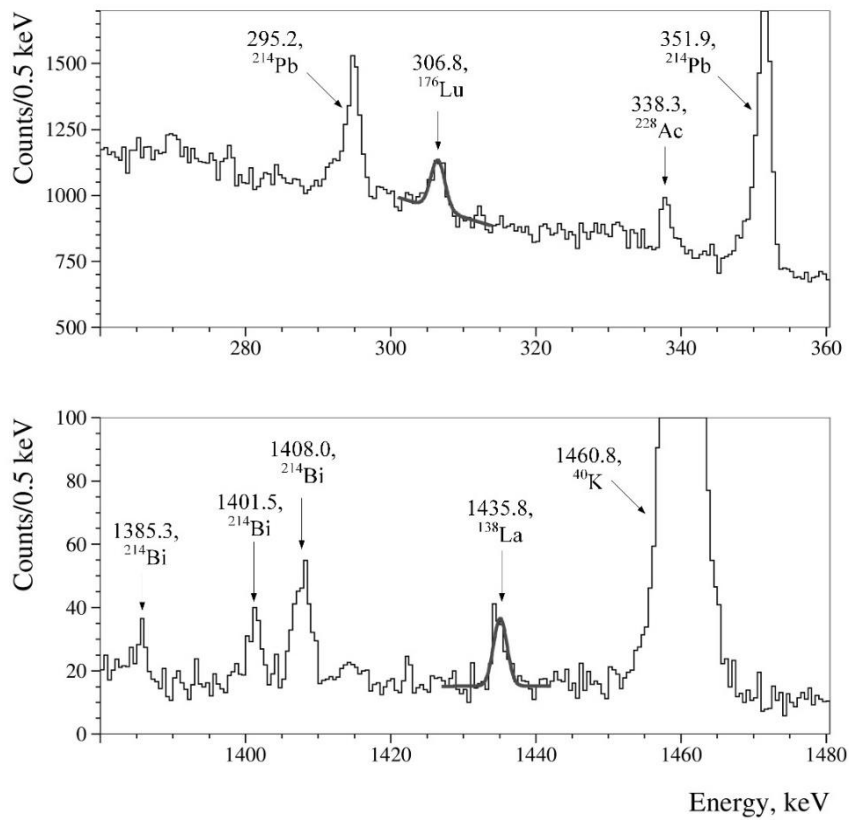

Fig. 3. Parts of the cumulative energy spectrum accumulated over 16375 h with the 2.381-kg Nd$_2$O$_3$ sample by the GeMulti detector in the energy regions of γ peaks 307 keV ($^{176}$Lu, *top*) and 1436 keV ($^{138}$La, *bottom*).

The peaks of $^{138}$La and $^{176}$Lu are observed also in the cumulative spectrum gathered with the GeMulti setup (see Fig. 3). Taking into account the areas of the peaks ($S_{307} = 919 \pm 112$ counts and $S_{1436} = 100 \pm 16$ counts) and the detection efficiencies (2.29 % and 1.24 % for 307 keV and 1436 keV, respectively, calculated with the help of the EGSnrc simulation package [29]), the activities of $^{138}$La and $^{176}$Lu in the sample are estimated as 0.057(9) and 0.29(4) mBq/kg, respectively.

Table 2. Radioactive contamination of the Nd$_2$O$_3$ before and after purification [22, 23] and the present study. Upper limits are given at 90 % C.L., the measured activities are given at 68 % C.L.

| Chain | Nuclei | Activity, mBq/kg | | |
|---|---|---|---|---|
| | | Before purification | After purification | Current measurements |
| | $^{40}$K | $16 \pm 8$ | $\leq 3.7$ | $\leq 1.8$ |
| | $^{137}$Cs | $\leq 0.80$ | $\leq 0.53$ | $\leq 0.04$ |
| | $^{176}$Lu | $1.1 \pm 0.4$ | $0.7 \pm 0.4$ | $0.29 \pm 0.04$ |
| | $^{138}$La | – | – | $0.057 \pm 0.009$ |
| $^{232}$Th | $^{228}$Ra | $\leq 2.1$ | $\leq 2.6$ | $\leq 0.3$ |
| | $^{228}$Th | $\leq 1.3$ | $\leq 1.0$ | $\leq 0.4$ |
| $^{235}$U | $^{235}$U | $\leq 1.7$ | $\leq 1.3$ | $\leq 1.3$ |
| $^{238}$U | $^{234}$Th | $\leq 28$ | $\leq 46$ | $\leq 5.4$ |
| | $^{226}$Ra | $15 \pm 0.8$ | $\leq 1.8$ | $\leq 1.9$ |

### 3.2. Two neutrino 2β decay of $^{150}$Nd to the $0_1^+$ level of $^{150}$Sm

Parts of the cumulative energy spectrum gathered with the Nd$_2$O$_3$ sample in the energy intervals 310 - 355 keV and 380 - 425 keV are shown in Fig. 4. One can see that there are no evident peaks with energies 334.0 and 406.5 keV in the experimental data. Thus, we can set only a lower limit on the half-life of $^{150}$Nd relatively to the 2β decay to the first $0^+$ excited level of $^{150}$Sm by using the following equation:

$$\lim T_{1/2} = \frac{\ln 2 \cdot \varepsilon \cdot N \cdot t}{\lim S}, \qquad (5)$$

where $\varepsilon$ is the full absorption peak detection efficiency of the 4 HP Ge detectors to the γ quanta with the energy of interest (calculated as 2.24% and 2.42 % for 334.0 and 406.5 keV, respectively, with the help of the EGSnrc simulation package [29]), $t$ is the time of measurements, $N$ is the number of $^{150}$Nd nuclei in the sample ($4.80 \times 10^{23}$), $\lim S$ is the number of events that can be excluded with a given confidence level (C.L.).

The values of lim $S$ were obtained from the fit of the experimental data in the energy intervals where the peaks are expected. The model of background in the energy interval of the 334.0 keV peak consists of a straight line (to describe continuous background), the peak searched for with energy 334.0 keV, and the gamma peaks due to the $^{228}$Ac (328.0 keV, 332.4 keV and 338.3 keV). The energy resolution of the peaks was bounded taking into account the dependence of the energy resolution on energy for γ quanta measured in the cumulative energy spectrum (see Sec. 2.2). The areas of the γ peaks of $^{228}$Ac were bounded according to their relative intensities (2.95 %, 0.4 % and 11.27 % for 328.0 keV, 332.4 keV and 338.3 keV, respectively), while the detection efficiency was assumed to be constant in the energy interval of the fit. The fit of the data in the energy interval 315 - 345 keV gives an area of the peak searched for 122 ± 76 counts (the result of the fit is shown in upper panel of Fig. 4), that is no evidence for the effect. A value of lim $S$ was estimated using the procedure proposed by Feldman and Cousins [30] as lim $S_{334}$ = 247 counts at 90 % C.L., which allowed to set a half-life limit $T_{1/2} \geq 5.6 \times 10^{19}$ y.

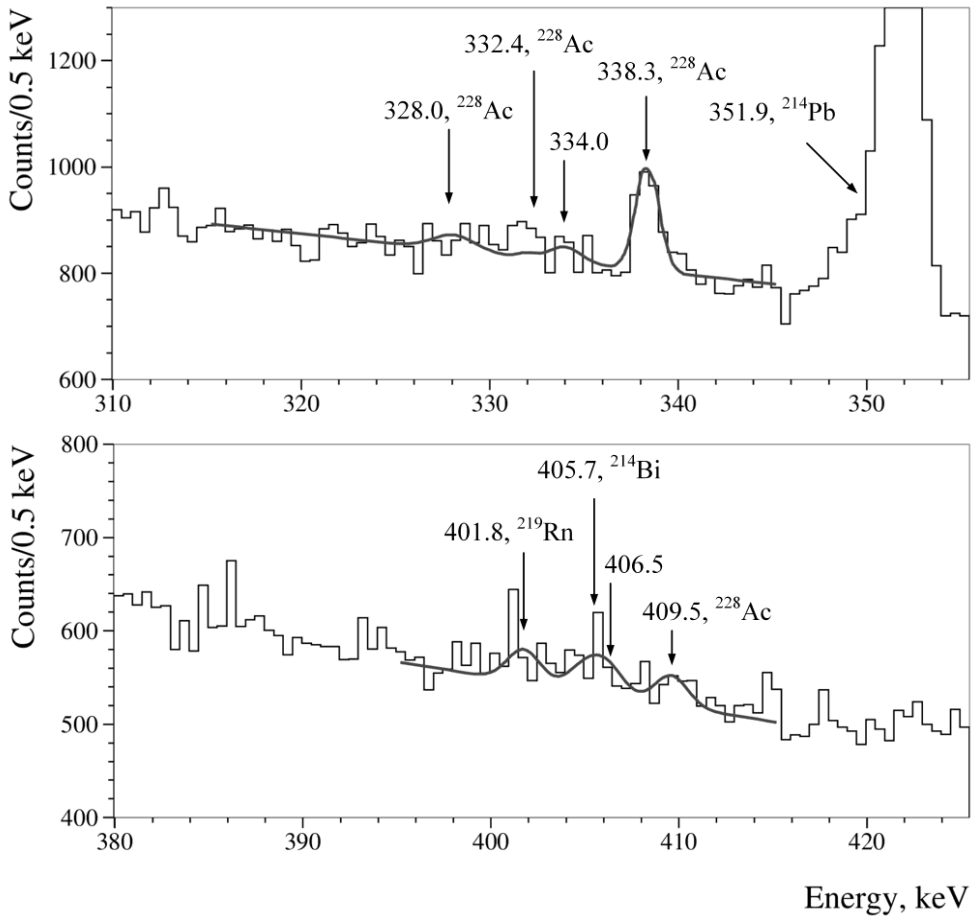

Fig. 4. The energy spectrum of the 2.381-kg Nd$_2$O$_3$ sample in the energy region of γ peaks 334.0 keV (*upper panel*) and 406.5 keV (*lower panel*). The fits of the data by the models of background (see text) are shown by solid lines. No evidence for the gamma's associated with the 2β decay of $^{150}$Nd to the $0^+_1$ 740.5 keV excited level of $^{150}$Sm have been observed.

A similar model was constructed to estimate lim $S$ for the peak expected at energy 406.5 keV. The model, in addition to a straight line and the peak searched for, included peaks of $^{219}$Rn (401.8 keV), $^{214}$Bi (405.7 keV) and $^{228}$Ac (409.5 keV). The areas of 405.7 keV and 409.5 keV peaks were bounded taking into account the areas of intensive peaks of $^{214}$Bi (609.3 keV) and $^{228}$Ac (338.3 keV, 911.2 keV), their relative intensities and the detection efficiencies. The fit of the energy spectrum in the energy interval 395 - 415 keV (lower panel in Fig. 4) provides an area of the effect searched for 78 ± 68 counts, that again gives no evidence of the effect. Using the recommendations in [30] one can obtain an excluded effect lim $S_{406} = 190$, which corresponds to a half-life of $T_{1/2} \geq 7.9 \times 10^{19}$ y at 90 % C.L.

A two-dimensional energy spectrum of coincidences between two detectors (events with a multiplicity 2) accumulated over 16375 h with the $Nd_2O_3$ sample is shown in Fig. 5 (*left panel*). By fixing the energy of events in one of the detectors to the energy of γ quantum that is expected to be in a cascade, a signal with energy corresponding to the other γ quanta in cascade are expected. An example of such coincidence is shown in Fig. 5 (*right panel*). The energy spectrum obtained in coincidence with the energy 609 ± 5 keV ($^{214}$Bi) in one of the detectors is shown in the right top panel. In the spectra there are peaks due to $^{214}$Bi with energies 768.4 keV, 1120.3 keV and 1238.1 keV. The energy spectrum accumulated in coincidence with energy 2615 ± 5 keV ($^{208}$Tl) is reported in right bottom panel. A gamma peak corresponding to the $^{208}$Tl decay with the energy 583.2 keV is clearly visible in the data.

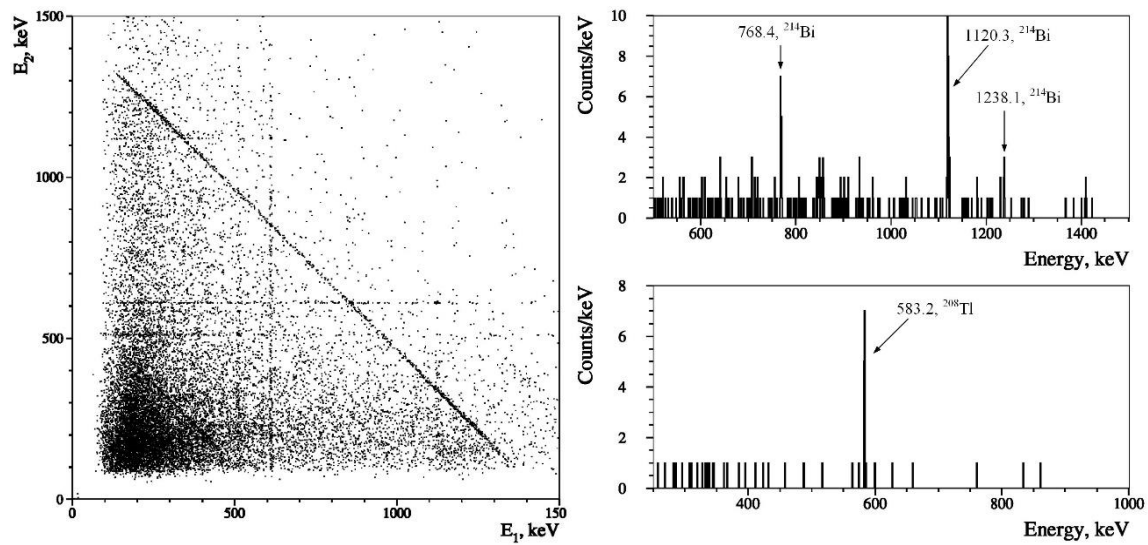

Fig. 5. The two-dimensional energy spectrum of events with multiplicity 2 accumulated in the coincidence mode (*left panel*). The coincidence spectra when the energy of one detector is fixed as (609 ± 5) keV ($^{214}$Bi, *top*) or (2615 ± 5) keV ($^{208}$Tl, *bottom*) (*right panel*). The spectra were obtained considering 16375 h of data gathered with the 2.381-kg $Nd_2O_3$ sample.

Fixing the energy of one of the detectors to the expected energy of γ quanta emitted in the 2β decay of $^{150}$Nd to the $0_1^+$ 740.5 keV excited level of $^{150}$Sm (334.0 keV or 406.5 keV, with the energy window ±1.4×FWHM), the coincidence signals at the supplemental energy (406.5 keV or 334.0 keV, respectively, see Fig. 6) have been observed. The area of each peak was estimated as $5.7^{+3.8}_{-2.6}$ counts (using the procedure proposed in [30]). Taking into account the detection efficiency calculated for this γ cascade ($4.3 \times 10^{-4}$) the obtained half-life of $^{150}$Nd to the $0^+_1$ 740.5 keV excited level of $^{150}$Sm is $T_{1/2} = 4.7^{+4.1}_{-1.9} \times 10^{19}$ y.

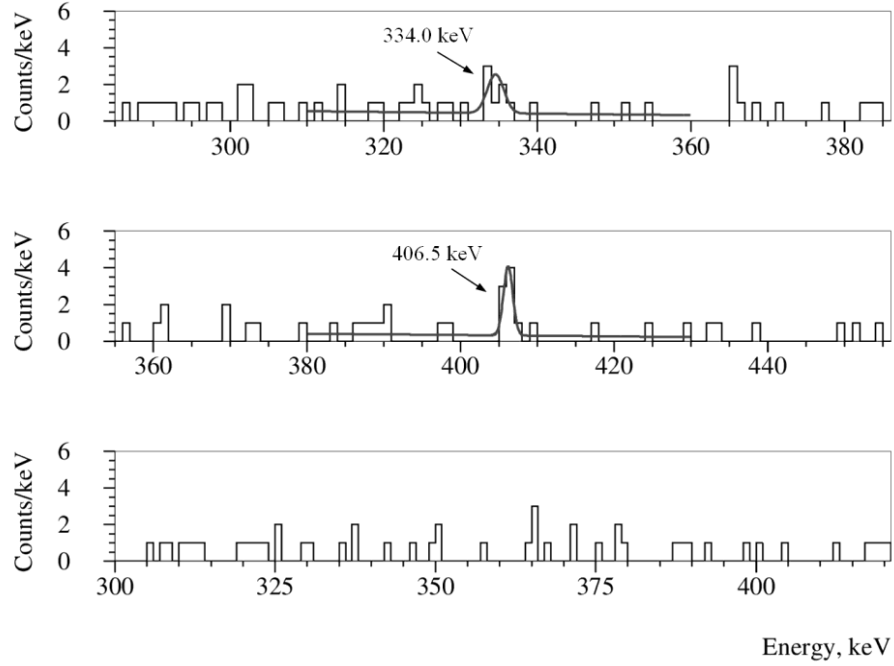

Fig. 6. The coincidence energy spectra accumulated over 16375 h by the GeMulti set-up with the 2.381-kg Nd$_2$O$_3$ sample, when the energy in one detector is fixed to the energy interval where γ quanta from the decay $^{150}$Nd → $^{150}$Sm ($0^+_1$, 740.5 keV): 406.5 keV ± 1.4×FWHM (*top*), 334.0 keV ± 1.4×FWHM (*middle*), are expected. The *bottom* spectrum shows a random coincidence background in the energy range of interest when energy of events in one of the detectors was taken as 375 keV ± 1.4×FWHM (no γ quanta with this energy are expected neither in the 2β decay of $^{150}$Nd nor in the decays of nuclides that are radioactive contamination of the Nd$_2$O$_3$ sample or the set-up).

The systematic uncertainties are due to the uncertainty of the Nd$_2$O$_3$ sample mass (0.04 %), the isotopic abundance of $^{150}$Nd in the sample (0.5 %), the live time (0.5 %), and the detection efficiency (10 %) [28]. Summing the systematic uncertainties in squares, one can obtain the following half-life of $^{150}$Nd relatively to the 2ν2β decay to the first $0^+$ 740.5 keV excited level of $^{150}$Sm:

$$T_{1/2} = [4.7^{+4.1}_{-1.9}(\text{stat}) \pm 0.5(\text{syst})] \times 10^{19} \text{ y} \qquad (6)$$

The half-life is in an agreement with the results of all the previous experiments (see Table 1 and Fig. 7).

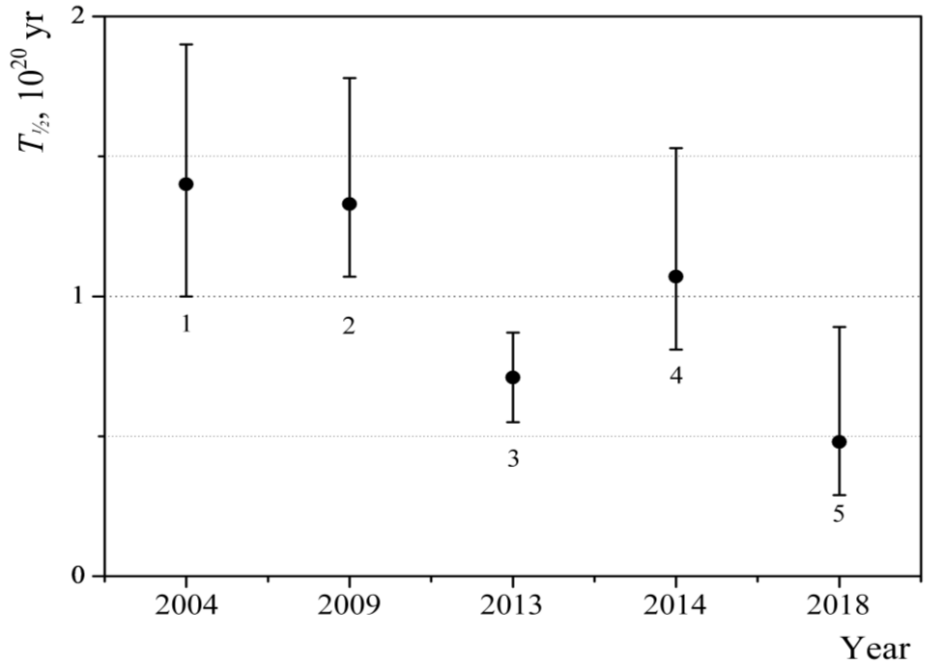

Fig. 7. The half-lives of $^{150}$Nd relatively to the two neutrino double beta decay transition to the first excited $0^+$ level of $^{150}$Sm measured in the experiment [18] (*1*), in the re-estimation of the experiment [18] in [19] (*2*), NEMO-3 experiment (preliminary result) [20] (*3*), measurements in the Kimballton Underground Research Facility [21] (*4*), current work (*5*).

## 4. Conclusions

Investigations of the double beta decay of $^{150}$Nd to the first $0^+$ 740.5 keV excited level of $^{150}$Sm are in progress at the Gran Sasso underground laboratory (Italy). The experiment utilizes a four-crystals ultra-low-background HP Ge spectrometer to detect γ quanta emitted in the cascade following the decay of $^{150}$Nd in a 2.381-kg sample of highly purified $Nd_2O_3$. In the data collected over 16375 h γ quanta with energies 334.0 keV and 406.5 keV are observed in coincidences between two detectors. The obtained half-life is $T_{1/2} = [4.7^{+4.1}_{-1.9}(\text{stat}) \pm 0.5(\text{syst})] \times 10^{19}$ y in an agreement with the results of previous experiments. The experiment is presently running to increase the statistics in order to improve the half-life value accuracy.

## References

1. V.I. Tretyak, Yu.G. Zdesenko, Tables of double beta decay data – an update. At. Data Nucl. Data Tables 80 (2002) 83.
2. R. Saakyan, Two-Neutrino Double-Beta Decay. Annu. Rev. Nucl. Part. Sci. 63 (2013) 503.
3. A.S. Barabash. Average and recommended half-life values for two-neutrino double beta decay. Nucl. Phys. A 935 (2015) 52.
4. J. Barea, J. Kotila, F. Iachello. Limits on Neutrino Masses from Neutrinoless

Double-β Decay. Phys. Rev. Lett. 109 (2012) 042501.

5. W. Rodejohann, Neutrinoless double-beta decay and neutrino physics. J. Phys. G 39 (2012) 124008.
6. F.F. Deppisch, M. Hirsch, H. Päs. Neutrinoless double-beta decay and physics beyond the standard model. J. Phys. G 39 (2012) 124007.
7. S.M. Bilenky, C. Giunti. Neutrinoless double-beta decay: A probe of physics beyond the Standard Model. Int. J. Mod. Phys. A 30 (2015) 1530001.
8. S. Dell'Oro et al. Neutrinoless Double Beta Decay: 2015 Review. AHEP 2016 (2016) 2162659.
9. J.D. Vergados, H. Ejiri, F. Šimkovic. Neutrinoless double beta decay and neutrino mass. Int. J. Mod. Phys. E 25 (2016) 1630007.
10. J. Schechter, J.W.F. Valle. Neutrinoless double-β decay in SU(2)×U(1) theories. Phys. Rev. D 25 (1982) 2951.
11. F. Vissani. Solar neutrino physics on the beginning of 2017. Nucl. Phys. At. Energy 18 (2017) 5.
12. V.S. Kolhinen et al. Double-β decay $Q$ value of $^{150}$Nd. Phys. Rev. C 82 (2010) 022501.
13. J. Meija et al. Isotopic compositions of the elements 2013 (IUPAC Technical Report), Pure Appl. Chem. 88 (2016) 293.
14. V. Artemiev et al. Half-life measurement of $^{150}$Nd 2β2ν decay in the time projection chamber experiment. Phys. Lett. B 345 (1995) 564.
15. A. De Silva et al. Double β decays of $^{100}$Mo and $^{150}$Nd. Phys. Rev. C 56 (1997) 2451.
16. R. Arnold et al. Measurement of the 2νββ decay half-life of $^{150}$Nd and a search for 0νββ decay processes with the full exposure from the NEMO-3 detector. Phys. Rev. D 94 (2016) 072003.
17. S.K. Basu, A.A. Sonzogni. Nuclear data sheets for A = 150. Nucl. Data Sheets 114 (2013) 435.
18. A.S. Barabash et al. Double-beta decay of $^{150}$Nd to the first $0^+$ excited state of $^{150}$Sm. JETP Lett. 79 (2004) 10.
19. A.S. Barabash et al. Investigation of ββ decay in $^{150}$Nd and $^{148}$Nd to the excited states of daughter nuclei. Phys. Rev. C 79 (2009) 045501.
20. S. Blondel. Optimisation du blindage contre les neutrons pour le démonstrateur de SuperNEMO et analyse de la double désintégration bêta du néodyme-150 vers les états excités du samarium-150 avec le détecteur NEMO-3. PhD thesis, LAL, Orsay, France, LAL 13-154 (2013).
21. M.F. Kidd et al. Two-neutrino double-β decay of $^{150}$Nd to excited final states in $^{150}$Sm. Phys. Rev. C 90 (2014) 055501.
22. O.G. Polischuk et al. Purification of lanthanides for double beta decay experiments. AIP Conf. Proc. 1549 (2013) 124.
23. R.S. Boiko. Chemical purification of lanthanides for low-background experiments. Int. J. Mod. Phys. A 32 (2017) 1743005.
24. N.A. Danilov et al. Exhaustive removal of thorium and uranium traces from neodymium by liquid extraction. Radiochem. 53 (2011) 269.

25. M. Laubenstein et al. Underground measurements of radioactivity. Appl. Radiat. Isotopes 61 (2004) 167.
26. V.I. Tretyak. TS2 – interactive system for one-dimensional spectra processing. Preprint KINR-90-35 (Kyiv, 1990).
27. R.B. Firestone et al. *Table of Isotopes.* 8th ed. (New York, 1996) and (CD update, 1998).
28. P. Belli et al. New observation of 2β2ν decay of $^{100}$Mo to the $0_1^+$ level of $^{100}$Ru in the ARMONIA experiment. Nucl. Phys. A 846 (2010) 143.
29. I. Kawrakow, D.W.O. Rogers, The EGSnrc code system: Monte Carlo simulation of electron and photon transport, NRCC Report PIRS-701, Ottawa, 2003.
30. G. Feldman, R. Cousins. Unified approach to the classical statistical analysis of small signals. Phys. Rev. D 57 (1998) 3873.